\documentclass[twocolumn,aps,amssymb,floatfix,prb,showpacs]{revtex4}
\usepackage{amsmath}
\usepackage{amsfonts}
\usepackage{graphicx}

\begin{document}

\title{Theory of the Bloch Oscillating Transistor}

\author{J. Hassel, and H. Sepp\"a}
\affiliation{VTT Information Technology, Microsensing, P.O. Box 1207,
FIN-02044 VTT}

\begin{abstract}
The Bloch oscillating transistor (BOT) is a device,
where single electron current through a normal tunnel junction can be used
to enhance Cooper pair current in a mesoscopic Josephson junction leading to
signal amplification. In this paper we develop a theory, where the BOT
dynamics is described as a two-level system. The theory is used to predict
current-voltage characteristics and small-signal response. Transition from
stable operation into hysteretic regime is studied. By identifying the
two-level switching noise as the main source of fluctuations, the
expressions for equivalent noise sources and the noise temperature are
derived. The validity of the model is tested by comparing the results with
simulations.
\end{abstract}

 \bigskip

\maketitle

\section{Introduction}

The Bloch oscillating transistor (BOT) \cite{sep1}$^-$\cite{has2} is a device
based on tuning the probability of interlevel switching in a mesoscopic
Josephson Junction\ (JJ). The equivalent circuit is shown in Fig. 1(a). The
current $I_{C}$ at the collector(C) -emitter(E) -circuit is controlled by the
base current $I_{B}$ leading to transistor-like operation. The physics behind
this is based on controlling the state of the JJ by means of quasiparticles
tunneling through the normal tunnel junction connected to the base electrode (B).

\begin{center}
\begin{figure}[!hbt]
\includegraphics[width=6cm]{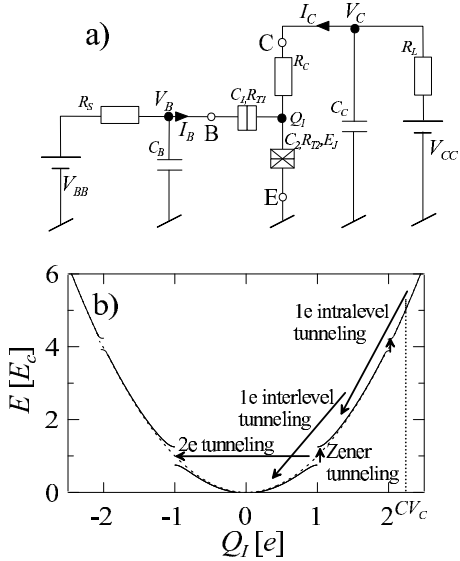}
\caption{(a) Schematic circuit of a BOT connected to a source and a load. Source
$R_{S}$ is connected to the base electrode B and load $R_{L}$ to the collector
electrode C. The lead capacitances from the electrodes to the ground are
$C_{B}$ and $C_{C}$. The BOT itself consists of\ a Josephson Junction (JJ)
between the emitter electrode E and the superconducting ''island'', normal
tunnel junction between B and the island and large $R\gg R_{Q}$ resistor
between C and the island. (b)\ The state diagram of the JJ and possible
transitions induced by charge tunneling through the junctions.}
\end{figure}
\end{center}

The state diagram as function of the ''island'' (quasi)charge $Q_{I}$ is shown
in Fig. 1(b)\cite{lik1}, where also the transitions are illustrated. It is
assumed that the Josephson coupling energy $E_{J}$ is smaller or of the same
order as the charging energy $E_{C}=e^{2}/2C_{\Sigma},$ and that
$R,R_{T1},R_{T2}\gtrsim R_{Q}$. Here $C_{\Sigma}$ is the total capacitance of
the island, $R$ is the collector resistor, and $R_{T1}$and $R_{T2}$ are the
tunnel resistances of the two junctions. The quantum resistance $R_{Q}%
=h/4e^{2}\approx6.5$ k$\Omega$. When C is biased at a point, where
$V_{C}\gtrsim e/C_{\Sigma}$, the charge tends to relax through the collector
resistor $R_{C}$ towards $V_{C}C_{\Sigma}$. Here $V_{C}$ is the collector
voltage. At $Q_{I}=e$ it is likely that a Cooper pair (CP) tunneling through
the JJ returns the island back to $Q_{I}=-e$. Repeating this cycle, the Bloch
Oscillation\cite{lik1}, leads to a net current through the CE-circuit. A
competing process is the Zener tunneling\cite{sch1}, which takes the system to
the upper bands ($\left|  Q_{I}\right|  >e$ in the extended band picture we
are using). There the Cooper pair tunneling is blocked. One or more
quasiparticles tunneling through the base junction may return the system back
to the lowest energy band i.e. back to $\left|  Q_{I}\right|  <e$.

The Zener tunneling probability between bands $n$ and $n-1$ is given as
$P_{n-1\rightleftharpoons n}^{Z}\approx\exp\left(  -\left(  \pi/8\right)
/n^{2n-1}\left(  E_{J}/E_{C}\right)  ^{2n}\left(  eE_{C}/\hbar I_{0}\right)
\right)  $. In the limit of small $E_{J}$ it approaches unity very rapidly as
$n$ increases. Therefore we can neglect Cooper pair tunneling at higher bands,
and consider the device as a two-level system. Below we will call the lowest
band with $n=0$ the ''first level'' and higher bands with $n\gtrsim1$ the
''second level''. Tunneling events, which cause transitions between the levels
will be called interlevel transitions (or upwards and downwards transitions)
and those, which cause transitions between higher bands, will be called
intralevel transitions. Controlling the probability of downwards transitions
leads to control of average $I_{C}$ and thus to transistor-like characteristics.

The BOT was recently experimentally realized \cite{del1}, and simulations
showed that its properties can be quantitatively predicted with a
computational model \cite{has2}. It is potentially useful in cryogenic
applications such as readout circuits of radiation detectors, or measurement
of small currents in quantum metrology. The aim of this article is to gain
more insight into the dynamics of the BOT\ and to find the noise properties.
To be able to do so, we derive an analytic theory, and study its applicability
by comparing the results to computational data. Small-signal parameters of
BOT\ as well as equivalent noise sources and the noise temperature are derived.

\section{Computational Model}

Since we assume that $E_{J}$ $\ll E_{C}$ we can interpret the quasicharge as
the real charge, and the energy diagram will reduce to the simple parabola
(dashed line in Fig. 1b). We can account for the Josephson coupling
perturbatively and calculate the CP tunneling probabilities using the
$P\left(  E\right)  $-theory \cite{dev1,ing1}. This takes the electromagnetic
environment into account. The consequence is that the tunneling does not
happen strictly at $Q_{I}=e$, but is rather represented with a finite
distribution\cite{has2}.

The simulation is done by integrating $Q_{I}$ in the time domain. The charge
$Q_{I}=Q_{2}-Q_{1}$ becomes%

\begin{equation}
\frac{dQ_{I}}{dt}=\frac{V_{C}-V_{2}}{R_{c}}-\left(  \frac{dQ_{I}}{dt}\right)
_{QP1}-\left(  \frac{dQ_{I}}{dt}\right)  _{QP2}-\left(  \frac{dQ_{I}}%
{dt}\right)  _{CP},\label{QI}%
\end{equation}

where the first term is the collector current, the second and the third terms
describe quasiparticle tunneling through the two junctions and the last term
describes the Cooper pair tunneling through the Josephson junction. The
corresponding tunneling rates as the function of the state of the system are
derived below. The voltage $V_{2}$ is across the JJ. The charges and voltages
across both junctions, $Q_{1}$ (base junction) and $Q_{2}$ (JJ)\ needed for
the tunneling rate computation below are obtained as follows:%

\begin{align}
Q_{1}  &  =\frac{1}{C_{\Sigma}}\left(  C_{1}C_{2}V_{B}-C_{1}Q_{I}\right) \\
Q_{2}  &  =\frac{1}{C_{\Sigma}}\left(  C_{1}C_{2}V_{B}+C_{2}Q_{I}\right)
\end{align}

The voltages across\ the junctions are $V_{i}=Q_{i}/C_{i}$. In numerical
simulations we will assume zero load resistance ($R_{L}=0$), i.e. the
collector voltage $V_{C}=$ $V_{CC}$ is fixed. In some simulations the base
electrode is assumed voltage biased (i.e. $R_{S}=0$ and $V_{B}=V_{BB}$). Note
that below we will also use parameter $V_{B}^{^{\prime}}=V_{B}-V_{C}$ as base
bias parameter for convenience. From the circuit point of view the choice
between $V_{B}$ and $V_{B}^{\prime}$ is the detailed bias arrangement.

In some simulations $R_{S}$ will be assumed much higher than the input
impedance. This fixes the base current $I_{B}$. In this case $V_{B}$ is
integrated as\ %

\begin{equation}
\frac{dV_{B}}{dt}=\frac{1}{C_{B}}\left(  I_{B}+\left(  \frac{dQ_{I}}%
{dt}\right)  _{QP1}\right)  ,
\end{equation}
where $I_{B}$ is the\ base current and $C_{B}$ is the cable capacitance. Here
we have assumed that $C_{B}\gg C_{\Sigma}$.

Both Cooper pair and quasiparticle tunneling probabilities are obtained from
the $P\left(  E\right)  $-theory \cite{ing1}, so that the effect of
electromagnetic environment (the collector resistor)\ is also included. The
given tunneling rates are towards the island. By reversing the signs of
voltages $V_{1}$ and $V_{2}$ the rates from the island are obtained. The
Cooper pair tunneling rate is%

\begin{equation}
\Gamma_{CP}\left(  V_{2}\right)  =\frac{\pi E_{J}^{2}}{2\hbar}P\left(
2eV_{2}\right)  ,\label{GCP}%
\end{equation}
and $P\left(  E\right)  $ is defined as%

\begin{equation}
P\left(  E\right)  =\frac{1}{2\pi\hbar}\int\limits_{-\infty}^{\infty}%
dt\exp\left(  J\left(  t\right)  +\frac{i}{\hbar}Et\right)  ,\label{pe}%
\end{equation}
where the phase correlation function $J\left(  t\right)  $ is%

\begin{align}
J\left(  t\right)    & =2\int\limits_{0}^{\infty}\frac{d\omega}{\omega}%
\frac{R_{C}/R_{Qi}}{1+\left(  \omega R_{C}C_{\Sigma}\right)  ^{2}}%
\label{Jt}\\
& \times\left[  \coth\left(  \frac{1}{2}\frac{\hbar\omega}{k_{B}T}\right)
\left(  \cos\left(  \omega t\right)  -1\right)  -i\sin\left(  \omega t\right)
\right]  . \nonumber
\end{align}
Here $R_{Q2}=h/4e^{2}\approx6.5$ k$\Omega$ is the quantum resistance for
Cooper pairs. For the quasiparticle tunneling formulas given below value
$R_{Q1}=h/e^{2}\approx25.9$ k$\Omega$ is used instead. Capacitance $C_{\Sigma
}=C_{1}+C_{2}$ is the total capacitance of both junctions, $T$ is the
temperature and $k_{B}$ the Bolzmann constant.

The quasiparticle tunneling rates through the junctions ($i=1$ for the base
junction and $i=2$ for the JJ) are%

\begin{align}
\Gamma_{QPi}\left(  V_{i}\right)    & =\frac{1}{e^{2}R_{Ti}}\int_{-\infty
}^{\infty}\int_{-\infty}^{\infty}dEdE^{\prime}\frac{N_{1}(E)}{N_{1}\left(
0\right)  }\frac{N_{2}(E^{^{\prime}}+eV_{i})}{N_{2}\left(  0\right)
}\label{Gqp}\\
& \times f\left(  E\right)  \left[  1-f\left(  E^{\prime}+eV_{i}\right)
\right]  P\left(  E-E^{\prime}\right)  ,\nonumber
\end{align}
where $R_{Ti}$ are the tunneling resistances. The densities of states on both
sides of the junctions are $N_{1}\left(  E\right)  $ and $N_{2}\left(
E\right)  $. For both NIS junction and the JJ $N_{1}\left(  E\right)
/N_{1}\left(  0\right)  =\Theta\left(  E^{2}-\Delta^{2}\right)  E/\sqrt
{E^{2}-\Delta^{2}}$. For the JJ $N_{2}\left(  E\right)  /N_{2}\left(
0\right)  =\Theta\left(  E^{2}-\Delta^{2}\right)  E/\sqrt{E^{2}-\Delta^{2}}$
and for the NIS Junction $N_{2}\left(  E\right)  /N_{2}\left(  0\right)  =1$.
The superconducting gap is $\Delta$, $f\left(  E\right)  $ is the Fermi
function and $\Theta\left(  E\right)  $ is the step function.

One should notice that the voltages appearing in the tunnel rate formulas are
time dependent as opposed to the standard $P\left(  E\right)  $-theory, where
the fixed bias voltage is used. By doing so we can account for the fact that
the tunneling probabilities depend on the state of the system.

\section{Analytic Theory}

In the theory derived below, BOT is modelled as a mapping of voltages $V_{B}$
and $V_{C}$ into currents $I_{C}$ and $I_{B}$. We assume that a single
tunneling event will not affect the voltages. This is the case, since
$C_{B},C_{C}\gg C_{\Sigma}$ in a practical experimental setup.

We assume that $1\ll E_{C}/kT\ll R_{C}/R_{Q}$ and $E_{J}\ll E_{C}$, which
means that the Cooper pair tunneling rate (Eq. (\ref{GCP})) reduces to a delta
spike centered at $\left|  Q_{I}\right|  =e$ \cite{has2}. This recovers our
interpretation of the two-level system with $\left|  Q_{I}\right|  <e$ as the
first level and $\left|  Q_{I}\right|  >e$ as the second level. We also assume
that $C_{2}\gg C_{1}$ and neglect quasiparticle tunneling through the JJ.
Below unnecessary subscripts for capacitances and charges are dropped, i.e.
$C\equiv C_{2}\equiv C_{\Sigma}$, $R_{T}\equiv R_{T1}$ and $Q\equiv
Q_{I}\equiv Q_{2}$. We analyze only the regime, where $V_{C}>e/C$ and
$V_{B}^{\prime}<0$, since this is interesting for the amplifier\ operation.

The collector current is written as%

\begin{equation}
I_{C}=\frac{1/\Gamma_{\uparrow}}{1/\Gamma_{\uparrow}+1/\Gamma_{\downarrow}%
}I_{S}-I_{B}\label{Icgen}%
\end{equation}
where the first term is the Cooper pair current through the Josephson junction
and $I_{B}$ is the single electron current through the base electrode. The
transition rates between the two levels are $\Gamma_{\uparrow} $ and
$\Gamma_{\downarrow}$. The ''saturation current'', i.e. current through the JJ
at the first level, is $I_{S}=2ef_{B}$, where $f_{B}$ is the Bloch
oscillations frequency.

The base current is%

\begin{equation}
I_{B}=-\frac{\left\langle N_{e}\right\rangle e}{1/\Gamma_{\uparrow}%
+1/\Gamma_{\downarrow}},\label{Ibgen}%
\end{equation}
where $\left\langle N_{e}\right\rangle $ is the number of electrons needed to
induce a downwards transition. Here we have neglected the possibility of
single-electron tunneling, when the system is at the first level. \ This is
justified, since typically the voltage $\left|  V_{1}\right|  $ is below the
gap voltage in that case. The Eqs. (\ref{Icgen}) and (\ref{Ibgen}) give
general IV characteristics for the BOT.

Between tunneling events $dQ/dt=\left(  V_{C}-Q/C\right)  /R$. By integrating
from $Q=-e$ to $Q=e$, i.e. over one Bloch period one gets%

\begin{equation}
f_{B}=\frac{1}{RC}\left[  \ln\left(  \frac{V_{C}/V_{Q}+1}{V_{C}/V_{Q}%
-1}\right)  \right]  ^{-1},
\end{equation}
or%

\begin{equation}
I_{S}=\frac{2e}{RC}\left[  \ln\left(  \frac{V_{C}/V_{Q}+1}{V_{C}/V_{Q}%
-1}\right)  \right]  ^{-1},\label{Is}%
\end{equation}
where we have defined $V_{Q}=e/C$.

The upwards tunneling rate (the Zener tunneling) can now be written as \cite{zai1}%

\begin{equation}
\Gamma_{\uparrow}=\frac{I_{S}}{2e\left\langle N\right\rangle },\label{Gup}%
\end{equation}
where%

\begin{equation}
\left\langle N\right\rangle =\exp\left(  \frac{I_{z}}{e}\frac{RC}{V_{C}%
/V_{Q}-1}\right)  -1\label{Nave}%
\end{equation}
is the average number of Cooper pairs in one sequence of Bloch oscillations.
One sequence here means the time between tunneling down to the first level and
tunneling back to the second level. The Zener avalanche current is $I_{z}=\pi
eE_{J}^{2}/8\hbar E_{c}$.

The downwards tunneling at low temperatures and for large $R$ is exclusively
due to single electron tunneling through the base junction. To calculate
$\left\langle N_{e}\right\rangle $ and $\Gamma_{\downarrow}$ we first derive
an approximation for $\Gamma_{QP1}$. The most general form is obtained from
Eq. (\ref{Gqp}). For many purposes, however, a piecewise linear approximation
is sufficient:%

\begin{align}
\Gamma_{QP1}\left(  t\right)   &  =\frac{1}{R_{T}C}\left(  \frac{V_{1}}{V_{Q}%
}-\frac{1}{2}\right)  \label{Gbase}\\
&  =\frac{1}{R_{T}C}\left(  \frac{Q\left(  t\right)  }{e}-\frac{V_{C}%
+V_{B}^{\prime}}{V_{Q}}-\frac{1}{2}\right)  ,\text{ if }\left|  V_{1}\right|
>V_{G}\nonumber\\
\Gamma_{QP1}\left(  t\right)   &  =0\text{, if }\left|  V_{1}\right|
<V_{G},\nonumber
\end{align}
where the gap-voltage is $V_{G}$, i.e., we neglect the leakage current at the
subgap voltages. In the most straightforward experimental realization the base
junction is a NIS\ junction. In this case $V_{G}=\Delta+V_{Q}/2$ including the
contribution of both superconducting and Coulomb gaps. In principle it is also
possible to realize the base junction as a NIN junction, i.e. with suppressed
superconductivity on the other electrode. Then the gap voltage is $V_{G}%
=V_{Q}/2$. In general, $\Gamma_{QP1}$ is a function of time due to the time
dependency of charge.

We now have to separate two different regimes to find analytic approximations
for $\left\langle N_{e}\right\rangle $ and $\Gamma_{\downarrow}$. If
$V_{C}<2V_{Q}$ one electron always suffices to return the system to the first
level, i.e., $\left\langle N_{e}\right\rangle =1$. In this case the
probability distribution of the first quasiparticle tunneling event after the
Zener tunneling is%

\begin{equation}
P\left(  t\right)  =\Gamma_{QP1}\left(  t\right)  \lim_{\Delta t\rightarrow
0}\prod_{j=1}^{t/\Delta t}\left(  1-\Gamma_{QP1}\left(  j\Delta t\right)
\right)  ,\label{Pdist}%
\end{equation}
which is the probability that an electron will tunnel at time $t$ times the
probability that it has not tunneled at earlier times. The charge in Eq.
(\ref{Gbase}) before the first quasiparticle event obeys simple $RC$%
-relaxation, i.e.%

\begin{equation}
Q\left(  t\right)  =CV_{C}\left(  1+\left(  \frac{V_{Q}}{V_{C}}-1\right)
e^{-\frac{1}{RC}t}\right)  ,\label{qzen}%
\end{equation}
where we assumed that Zener tunneling occurs at $t=0$ (or equivalently that
$Q\left(  0\right)  =e$). The average rate $\Gamma_{\downarrow}$ is the
inverse of the weight of the distribution given by Eq. (\ref{Pdist}). Thus%

\begin{align}
\left\langle N_{e}\right\rangle  &  =1\label{ne1}\\
\Gamma_{\downarrow} &  =\frac{1}{\int_{0}^{\infty}tP\left(  t\right)
dt}.\label{gdown1}%
\end{align}

In general, the downwards rate has to be evaluated numerically from
(\ref{gdown1}). However, if we further assume that the transient in Eq.
(\ref{qzen}) is short, Eq. (\ref{Pdist}) reduces to a simple exponential
distribution. This is equivalent to assuming that $Q\left(  t\right)  \approx
CV_{C}$ at all times, and thus Eq. (\ref{Gbase}) also becomes time
independent. Now simply $\Gamma_{\downarrow}=\Gamma_{QP1}$, and it follows%

\begin{equation}
\Gamma_{\downarrow}=-\frac{1}{CR_{T}}\left(  \frac{V_{B}^{\prime}}{V_{Q}%
}+\frac{1}{2}\right)  .\label{gdown2}%
\end{equation}

If $V_{C}>2V_{Q}\,$it is possible that the first electron tunneling through
the base junction does not cause a transition to the first level, but some of
the single quasiparticle events lead to intralevel transitions instead. This
was found to have a dramatic effect in the experiments\cite{del1},\cite{has2},
and is found to be an important issue from the device optimization point of
view as well. To solve $\left\langle N_{e}\right\rangle $ and $\Gamma
_{\downarrow}$ analytically from Eqs. (\ref{Gbase}) and (\ref{qzen}) in this
case is unfortunately impossible. To find a sufficient approximation for our
purposes, we have solved the problem numerically and searched for a proper
fitting function. For simplicity we have assumed an NIN junction at the base
electrode, i.e. $V_{G}=V_{Q}/2$ in Eq. (\ref{Gbase}). Some fits are shown in
Fig. 2 and the result is%

\begin{align}
\left\langle N_{e}\right\rangle  &  =0.04\left(  \frac{R_{T}}{R}\right)
^{2}\label{ne3}\\
&  \times\exp\left(  0.3\exp\left(  1.8\frac{V_{C}}{V_{Q}}+0.27\frac
{V_{C}V_{B}^{\prime}}{V_{Q}^{2}}-0.2\frac{V_{B}^{\prime}}{V_{Q}}\right)
\right)  +1 \nonumber \\
\Gamma_{\downarrow}^{-1} &  =1.2e\frac{R+R_{T}}{V_{B}^{\prime}}\left(
1-\left\langle N_{e}\right\rangle \right)  \label{gdown3}\\
&  +RC\left(  2.5\frac{R_{T}}{R}+1.1\right)  \left(  \frac{V_{Q}}%
{V_{B}^{\prime}}\right)  ^{2}.\nonumber
\end{align}
The fit is accurate, when $R_{T}\lesssim R$. The weaker dependence indicated
by the unity term in Eq. (\ref{ne3}) and $\left(  2.5R_{T}/R+1.1\right)
\left(  V_{B}^{\prime}/V_{Q}\right)  ^{2}$ term in Eq. (\ref{gdown3}) dominate
at $V_{C}\approx2V_{Q}$ and large $\left|  V_{B}^{\prime}\right|  $. In this
case only one quasiparticle is needed to induce a downwards transition. This
is possible, if the tunneling occurs during the transient immediately after
the Zener tunneling, while still $Q\left(  t\right)  <2e$. The terms are
actually an approximation of Eqs. (\ref{ne1}) and (\ref{gdown1}) at
$V_{C}\approx2V_{Q}$. The $\exp(0.3\exp(...))$-term dominates, when several
tunneling events are needed to induce an interlevel transition. The very
strong dependence is roughly explained as follows. Let us assume that
$2V_{Q}<V_{C}<3V_{Q}$ and the island charge is initially $Q\approx CV_{C}$
(see Fig. 1). Now at least two quasiparticles tunneling rapidly one after
another are needed to induce a downwards transition. The quasiparticle
tunneling probability according to Eq. (\ref{Gbase}) is at its maximum, when
$Q\approx CV_{C}$. However, after the first tunneling event $Q$ drops down to
$CV_{C}-e$ and therefore the probability also drops. Hence the probability for
the second quasiparticle to tunnel before the charge relaxes back to $Q>2e$ is
small. The charge therefore tends to oscillate between $Q\approx CV_{C}$ and
$Q\approx CV_{C}-e$ for a long time before the rather improbable event at
$Q<2e$ happens. This generates a large quantity of intralevel transitions thus
increases $\left\langle N_{e}\right\rangle $ and $\Gamma_{\downarrow}$.

\begin{center}
\begin{figure}[!b]
\includegraphics[width=8cm]{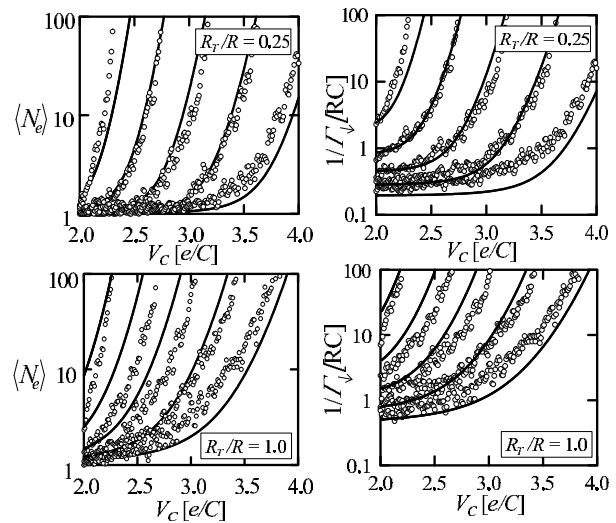}
\caption{Computed data for $\left\langle N_{e}\right\rangle $ and $\Gamma
_{\downarrow}$ obtained by solving Eqs. (\ref{Gbase}) and (\ref{qzen})
numerically (open circles). In each frame the base voltage is varied as
$V_{B}^{\prime}=-$1.0$e/C,$ -1.5$e/C,$ -2.0$e/C,$ -2.5$e/C$, -3.0$e/C$ from
left to right. The lines correspond the fits, i.e. Eqs. (\ref{ne3}) and
(\ref{gdown3}).}
\end{figure}
\end{center}

\section{Comparing Numeric and Analytic IV\ curves}

In Fig. 3(a) we show a simulated set of $I_{C}-V_{C}$ curves (open circles),
where the base is voltage biased. The base voltage $V_{B}^{\prime}$ is varied,
while other parameters are $R=1.5$ M$\Omega$, $C=0.2$ fF, $R_{T}=12$ M$\Omega
$, $E_{J}/E_{C}=0.1$, $T=40$ mK and $\Delta=1.5$ mV. Corresponding analytic
curves (solid lines) are calculated from Eq. (\ref{Icgen}) using the
approximation of Eqs. (\ref{ne1}) and (\ref{gdown1}) when calculating
$\left\langle N_{e}\right\rangle $ and $\Gamma_{\downarrow}$. The agreement is
reasonably good. An error is introduced due to the assumptions made on the
shape of $\Gamma_{QP1}\left(  V_{1}\right)  $. If one uses the full
temperature-dependent form (Eq. (\ref{Gqp})) instead of Eq. (\ref{Gbase}) the
agreement is improved especially at low values of $V_{B}^{\prime}$ as denoted
by the dashed line in Fig. 3.

\begin{center}
\begin{figure}[!t]
\includegraphics[width=7cm]{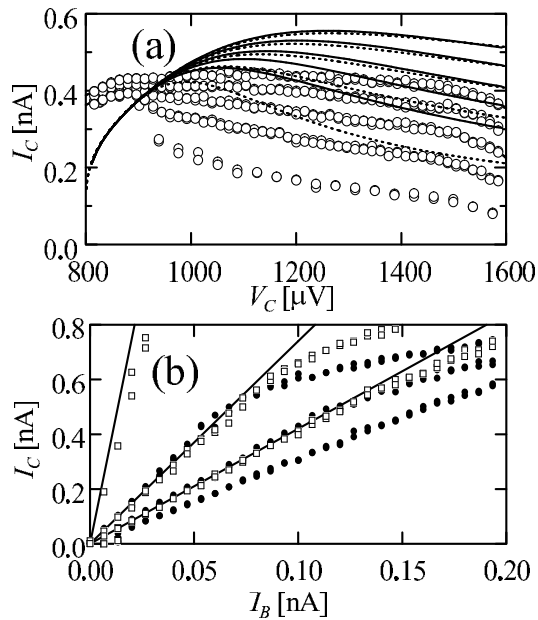}
\caption{(a)\ Computed $I_{C}-V_{C}$ plots with $R=1.5$ M$\Omega$, $C=0.2$ fF,
$R_{T}=12$ M$\Omega$, $E_{J}/E_{C}=0.1$, $T=40$ mK and $\Delta=1.5$ mV (open
circles). The base voltage has been varied as $V_{B}^{\prime}=-$2.5$e/C,$
-3.0$e/C,$ -3.5$e/C,$ -4.0$e/C$, -4.5$e/C$ from down to top. Solid lines
represent analytic values calculated from Eq. (\ref{Icgen}) together with
approximations from Eqs. (\ref{ne1}) and (\ref{gdown1}). Dashed lines are
corrected analytic curves, which take base junction nonlinearity at the finite
temperature into account. (b) Computed $I_{C}-I_{B}$ plots for the same device
(solid circles) at $V_{C}=$1.25$e/C,$ 1.5$e/C,$ 1.75$e/C$ from up to down. The
open squares shows the same simulation without ''Cooper pair back-tunneling''
and lines show analytic predictions.}
\end{figure}
\end{center}

The remaining disagreement can be found to be related to the temperature
dependence of Cooper pair tunneling probabilities. Even if $E_{C}/kT$ is as
high as about 120, incoherent Cooper pair tunneling enhances Cooper pair
current at $V_{C}\approx V_{Q}=800$ $\mu$V. The lower value of simulated
$I_{C}$ at larger values of $V_{C}$ was found to be due to the fact that after
a Cooper pair tunnels to the island it can immediately tunnel out of the
island due to incoherent Cooper pair tunneling. This effectively suppresses
$\left\langle N\right\rangle ,$ or equivalently enhances $\Gamma_{\uparrow}$.
The effect is especially visible in Fig. 3(b), where a set of simulations with
a current biased base electrode is performed for the same device. The
simulated curves (solid circles)\ fall below the theoretical curves (lines)
$I_{C}=\left(  2\left\langle N\right\rangle +1\right)  I_{B}$ (see also
Section VI), i.e. the current gain is suppressed. However, if we artificially
forbid the ''Cooper-pair back-tunneling'' in the simulation (open squares in
Fig. 3(b)) the agreement is clearly improved. This shows that the effect
indeed is the main factor suppressing the current gain in the mode of
operation governed by approximation given in Eqs. (\ref{ne1}) and
(\ref{gdown1}). Another mechanism due to spontaneous downwards transitions was
discussed in Ref. \cite{del2}, but it was found to be insignificant in this case.

\begin{center}
\begin{figure}[!b]
\includegraphics[width=8cm]{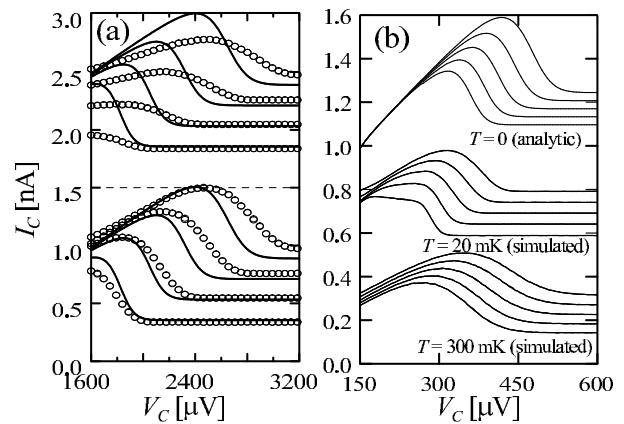}
\caption{(a) Computed $I_{C}-V_{C}$ plots for a device otherwise similar to that of
Fig. 3 except $R_{T}=375$ k$\Omega$, $E_{J}/E_{C}=0.2$ and $\Delta=0$ (an NIN
base junction). The base voltages are $V_{B}^{\prime}=-$1.0$e/C,$ -1.5$e/C,$
-2.0$e/C,$ -2.5$e/C$ from down to top (open circles) . Analytic IV curves
(solid lines)\ are calculated from (\ref{Icgen}) together with approximations
from Eqs. (\ref{ne3}) and (\ref{gdown3}). The upper set (lifted by 1.5 nA for
clarity) shows the result with the full simulation model, while the lower set
shows the result without ''Cooper pair back-tunneling''. (b)\ Analytic and
computed $I_{C}-V_{C}$ plots for a device having $R=500$ k$\Omega$, $C=1.2$
fF, $R_{T}=250$ k$\Omega$, $\Delta=400$ $\mu V$ and $E_{J}/E_{C}=0.3$. The two
topmost sets have been lifted by 0.5 nA and 0.8 nA for clarity.}
\end{figure}
\end{center}

As the tunnel resistance of the base electrode is decreased and Josephson
coupling increased in simulations and experiments\cite{del1},\cite{has2} it
has been found that the active bias region moves towards higher $V_{C}$
indicating that the approximation of $\left\langle N_{e}\right\rangle $ and
$\Gamma_{\downarrow}$ given in\ Eq:s (\ref{ne3}) and (\ref{gdown3}) becomes
relevant. In Fig. 4(a) a set of simulations with parameters similar to those
considered above, with exceptions $R_{T}=375$ k$\Omega$, $E_{J}/E_{C}=0.2$ and
$\Delta=0$ for the base junction (i.e. we have assumed that the base junction
is a NIN junction here). At the upper set it is again shown a set of simulated
and analytic $I_{C}-V_{C}$ curves showing a reasonable agreement. The
agreement is again further improved by forbidding the ''Cooper-pair
back-tunneling'' in the simulation, which is shown in the lower set of curves.

The devices analyzed above have relatively small $C$, which makes the voltages
at which they are operated (of order $V_{Q}$) rather high. The advantage is
that the temperature dependence is minimized as $E_{C}/kT$ increases. The
drawback is that higher band-gap materials are needed, since the voltage
across the Josephson junction must be below $2\Delta$. The above devices could
be realized using Nb technology, for which 2$\Delta\approx3$ mV. More
conventional Al-junctions have 2$\Delta\approx400$ $\mu$V, whence capacitances
have to be above\ or around 1 fF. In Fig. 4(b) a set of $I_{C}-V_{C}$ curves
are shown for a device with $R=500$ k$\Omega$, $C=1.2$ fF, $R_{T}=250$
k$\Omega$, $E_{J}/E_{C}=0.3$. The topmost set consists of analytic curves,
where at $V_{C}\lesssim2V_{Q}\approx270$ $\mu$V approximation of Eqs.
(\ref{ne1}) and (\ref{gdown1}) and at $V_{C}\gtrsim2V_{Q}$ approximation of
Eqs. (\ref{ne3}) and (\ref{gdown3}) is used. The two lower sets are simulated
at $T=20$ mK and $T=300$ mK. Although again qualitatively similar, at $T=20$
mK the main source of disagreement is the enhancement of $\Gamma_{\uparrow}$
at a finite temperature. At $T=300$ mK the spike is spread, since at
relatively large temperatures (now $E_{C}/kT\approx2.6$) also $\Gamma
_{\downarrow}$ is increased due to incoherent Cooper pair tunneling in a same
sense as indicated in\ Ref. \cite{del2}.

\section{Small signal parameters}

A complete small-signal model for the BOT\ can be given as conductance matrix%

\begin{equation}
\left[
\begin{array}
[c]{c}%
i_{C}\\
i_{B}%
\end{array}
\right]  =\left[
\begin{array}
[c]{cc}%
G_{out} & g_{m}\\
g_{x} & G_{in}%
\end{array}
\right]  \left[
\begin{array}
[c]{c}%
v_{C}\\
v_{B}%
\end{array}
\right]  ,\label{condmat}%
\end{equation}
where $i_{C},i_{B},v_{C},v_{B}$ are the small-signal components of collector
and base currents and voltages, i.e. small variations around the point of
operation. The definitions of small-signal parameters are listed below:%

\begin{align}
G_{in}  &  =\left(  \frac{\partial I_{B}}{\partial V_{B}}\right)  _{V_{C}}\\
g_{m}  &  =\left(  \frac{\partial I_{C}}{\partial V_{B}}\right)  _{V_{C}}\\
g_{x}  &  =\left(  \frac{\partial I_{B}}{\partial V_{C}}\right)  _{V_{B}}\\
G_{out}  &  =\left(  \frac{\partial I_{C}}{\partial V_{C}}\right)  _{V_{B}}.
\end{align}
Note that $V_{B}$ is kept constant in the last two lines. This is the natural
choice, if the circuit shown in Fig. 1(a)\ is used. However, if the emitter is
voltage biased instead of the collector, $V_{B}^{\prime}$ should be fixed
instead. The choice does not have an effect on the analysis below, since we
will mostly be assuming small $R_{L}$, whence $V_{C}$ is constant. For
completeness, the general formulas for small signal parameters are given here anyway.

Using Eqs. (\ref{Icgen}) and (\ref{Ibgen}) we can rewrite the parameters as%
\begin{align}
G_{in} &  =e\left\langle N_{_{e}}\right\rangle \left(  \frac{\Gamma_{\uparrow
}}{\Gamma_{\uparrow}+\Gamma_{\downarrow}}\right)  ^{2}\frac{\partial
\Gamma_{\downarrow}}{\partial V_{B}}\left(  \beta_{B}-1\right)  \label{Gingen}%
\\
g_{m} &  =I_{S}\frac{\Gamma_{\uparrow}}{\left(  \Gamma_{\uparrow}%
+\Gamma_{\downarrow}\right)  ^{2}}\frac{\partial\Gamma_{\downarrow}}{\partial
V_{B}}+G_{in}\label{gmgen}\\
g_{x} &  =-e\left\langle N_{e}\right\rangle \frac{\Gamma_{\downarrow}%
\Gamma_{\uparrow}}{\Gamma_{\uparrow}+\Gamma_{\downarrow}}(\frac{1}%
{\left\langle N_{e}\right\rangle }\frac{\partial\left\langle N_{e}%
\right\rangle }{\partial V_{C}}+\label{gxgen}\\
&  +\frac{\Gamma_{\downarrow}}{\Gamma_{\uparrow}\left(  \Gamma_{\uparrow
}+\Gamma_{\downarrow}\right)  }\frac{\partial\Gamma_{\uparrow}}{\partial
V_{C}}+\frac{\Gamma_{\uparrow}}{\Gamma_{\downarrow}\left(  \Gamma_{\uparrow
}+\Gamma_{\downarrow}\right)  }\frac{\partial\Gamma_{\downarrow}}{\partial
V_{C}})\nonumber\\
G_{out} &  =I_{S}\frac{\Gamma_{\uparrow}\Gamma_{\downarrow}}{\Gamma_{\uparrow
}+\Gamma_{\downarrow}}(\frac{1}{\Gamma_{\downarrow}\left(  \Gamma_{\uparrow
}+\Gamma_{\downarrow}\right)  }\frac{\partial\Gamma_{\downarrow}}{\partial
V_{C}}+\label{Goutgen}\\
&  -\frac{1}{\Gamma_{\uparrow}\left(  \Gamma_{\uparrow}+\Gamma_{\downarrow
}\right)  }\frac{\partial\Gamma_{\uparrow}}{\partial V_{C}}+\frac{1}%
{\Gamma_{\uparrow}}\frac{\partial I_{S}}{\partial V_{C}})+g_{x},\nonumber
\end{align}
where the fact that $\Gamma_{\uparrow}$, $\left\langle N\right\rangle $ and
$I_{S}$ are independent of $V_{B}$ is used. Indices for constant quantities
are dropped here for clarity. We have also defined%

\begin{equation}
\beta_{B}=-\frac{\Gamma_{\downarrow}\left(  \Gamma_{\uparrow}+\Gamma
_{\downarrow}\right)  }{\Gamma_{\uparrow}\left\langle N_{_{e}}\right\rangle
}\left(  \frac{\partial\left\langle N_{_{e}}\right\rangle }{\partial V_{B}%
}/\frac{\partial\Gamma_{\downarrow}}{\partial V_{B}}\right) \label{betabdef}%
\end{equation}
In the approximation of Eq. (\ref{ne1}) $\beta_{B}$ is zero,
since$\ \left\langle N_{_{e}}\right\rangle $ is constant. Using Eqs.
(\ref{ne3}) and (\ref{gdown3}) instead makes values $\beta_{B}\approx1$
possible. We call $\beta_{B}$ the ''hysteresis parameter'' of the BOT.

For some purposes it is also useful to define current gain%

\begin{equation}
\beta=-\left(  \frac{\partial I_{C}}{\partial I_{B}}\right)  _{V_{C}}%
=-\frac{g_{m}}{G_{in}}.\label{betadef}%
\end{equation}
Using Eqs. (\ref{Gingen}) and (\ref{gmgen}) this is given as%

\begin{equation}
\beta=\frac{1}{e}\frac{I_{S}}{\Gamma_{\uparrow}\left\langle N_{_{e}%
}\right\rangle \left(  1-\beta_{B}\right)  }+1.\label{betagen}%
\end{equation}

\section{Amplifier properties}

In this section we assume that $R_{L}\ll1/\left|  G_{out}\right|  $, i.e. that
BOT is read out with a current amplifier and thus $V_{C}$ is constant. For
example, if one uses a dc SQUID as a postamplifier, $R_{L}$ is close to zero.
Even though a voltage amplifier is used $R_{L}\approx0$ can be realized in
practice using current feedback. We will discuss two limits, one with
approximation given in Eqs. (\ref{ne1}) and (\ref{gdown2}) for evaluating
$\Gamma_{\downarrow}$ and $\left\langle N_{e}\right\rangle $. In this case
$\beta_{B}=0$. The second approximation uses Eqs. (\ref{ne3}) and
(\ref{gdown3}) to evaluate $\Gamma_{\downarrow}$ and $\left\langle
N_{e}\right\rangle $. Then it is possible to tune $\beta_{B}$ close to unity.
The emphasis is to find noise properties of the BOT\ at low frequencies.

The noise current at the output of the BOT\ is obtained by assuming that the
dominant noise mechanism is the two-level switching noise. At low frequencies
the corresponding sperctral noise density of the output current fluctuations
is (see e.g. \cite{kog1})%

\begin{equation}
S_{i,out}=4I_{S}^{2}\frac{\Gamma_{\downarrow}\Gamma_{\uparrow}}{\left(
\Gamma_{\uparrow}+\Gamma_{\downarrow}\right)  ^{3}},\label{Sout}%
\end{equation}
where the collector current switches between values $I_{C}=0$ and $I_{C}%
=I_{S}$. In other words we have neglected the single electron leakage from the
base electrode here.

\begin{center}
\begin{figure}[!b]
\includegraphics[width=8cm]{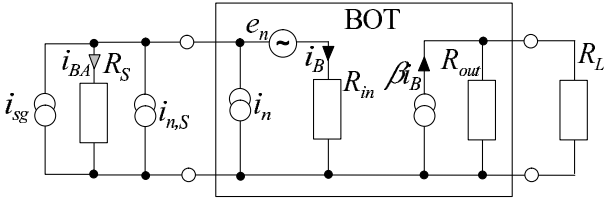}
\caption{A graphical representation of the small signal model of BOT in the limit of
small $R_{L}$. The noise added by BOT is represented with equivalent sources
$i_{n}$ end $e_{n}$.}
\end{figure}
\end{center}

The small-signal noise model for the BOT in the limit of small $R_{L}$ is
shown in Fig. 5. The signal and the noise from the source are described as
current generators $i_{sg}$ and $i_{n,S}$ in parallel with the source
resistance $R_{S}$. The input and output impedances are $R_{in}=1/G_{in}$ and
$R_{out}=1/G_{out}$. The current generator $\beta i_{B}$ at the output
accounts for the gain. The noise added by the BOT is represented in a standard
fashion (see e.g. \cite{eng1}) by equivalent voltage and current noise
generators ($e_{n}$ and $i_{n}$, respectively) at the input. According to Fig.
5 the output noise of the BOT excluding the contribution of the source
($i_{n,S}=0$) at the output is%

\begin{equation}
S_{i,out}^{1/2}=\frac{1}{R_{in}+R_{S}}\beta S_{en}^{1/2}+\frac{1/R_{in}%
}{1/R_{in}+1/R_{S}}\beta S_{in}^{1/2},\label{Sinout}%
\end{equation}
where $S_{en}$ and $S_{in}$ are the spectral density functions corresponding
to $e_{n}$ and $i_{n}$, respectively. Note that $e_{n}$ and $i_{n}$ and are
fully correlated with equal phases. Choosing%

\begin{align}
S_{en}^{1/2} &  =\frac{2I_{S}}{-g_{m}}\sqrt{\frac{\Gamma_{\downarrow}%
\Gamma_{\uparrow}}{\left(  \Gamma_{\uparrow}+\Gamma_{\downarrow}\right)  ^{3}%
}}\label{Sen}\\
S_{in}^{1/2} &  =\frac{2I_{S}}{\beta}\sqrt{\frac{\Gamma_{\downarrow}%
\Gamma_{\uparrow}}{\left(  \Gamma_{\uparrow}+\Gamma_{\downarrow}\right)  ^{3}%
}},\label{Sin}%
\end{align}
the generators become independent of $R_{S}$ and produce the output noise in
Eq. (\ref{Sout}) correctly. However, the backaction noise (i.e. the noise
current $i_{BA}$ through or voltage accross $R_{S}$) is not correctly
predicted by the model.

The noise figure, defined as the ratio of total noise at the output divided by
the noise contributed by the BOT, is%

\begin{equation}
F=1+S_{i,out}\left[  \left(  \frac{\beta R_{S}}{R_{in}+R_{S}}\right)
^{2}S_{in,S}\right]  ^{-1},\label{nfig}%
\end{equation}
where $S_{in,S}=4kT_{0}/R_{S}$ is the spectal density function of $i_{n,S}$
and $T_{0}$ is a reference temperature. One gets optimum impedance $R_{opt}$
and corresponding minimum noise temperature $T_{n}$ by minimizing $F$ and
using the definiton $F=1+T_{n}/T_{0}$. It follows%

\begin{align}
R_{opt} &  =\sqrt{\frac{S_{en}}{S_{in}}}=\left|  R_{in}\right| \label{Roptdef}%
\\
T_{n} &  =\frac{1}{k_{B}}\sqrt{S_{en}S_{in}}=\frac{\left|  R_{in}\right|
S_{in}}{k_{B}}.\label{Tndef}%
\end{align}
The correlation of the two sources shows in Eq. (\ref{Tndef}) in such a way
that the prefactor is $1/k_{B}$ instead of $1/2k_{B}$, which is the case for
uncorrelated sources. The difference stems from the fact that now the
amplitudes of the two sources rather than the powers are summed.

If the approximation of Eqs. (\ref{ne1}) and (\ref{gdown2}) is used to
evaluate $\left\langle N_{e}\right\rangle $ and $\Gamma_{\downarrow}$ (whence
also $\beta_{B}=0)$, one gets for gain and noise parameters%

\begin{align}
\beta &  =2\left\langle N\right\rangle +1\label{betaap0}\\
g_{m} &  =-\frac{2\left\langle N\right\rangle +1}{R_{T}}\frac{1}{\left(
1+\Gamma_{\downarrow}/\Gamma_{\uparrow}\right)  ^{2}}\\
S_{in}^{1/2} &  =\sqrt{-\frac{4e}{R_{T}}\left(  V_{B}^{\prime}+\frac{V_{Q}}%
{2}\right)  \left(  1+\frac{\Gamma_{\downarrow}}{\Gamma_{\uparrow}}\right)
^{-3}}\\
S_{en}^{1/2} &  =\sqrt{-4eR_{T}\left(  V_{B}^{\prime}+\frac{V_{Q}}{2}\right)
\left(  1+\frac{\Gamma_{\downarrow}}{\Gamma_{\uparrow}}\right)  }\\
R_{opt} &  =R_{T}\left(  1+\Gamma_{\downarrow}/\Gamma_{\uparrow}\right)
^{2}\label{tnap0}\\
T_{n} &  =-\frac{4e}{k_{B}}\left(  V_{B}^{\prime}+\frac{V_{Q}}{2}\right)
\left(  1+\frac{\Gamma_{\downarrow}}{\Gamma_{\uparrow}}\right)  ^{-1}.
\end{align}
In this mode the BOT acts as a simple ''charge multiplier'', where one
electron trigs $\left\langle N\right\rangle $ Cooper pairs, thus
$\beta=2\left\langle N\right\rangle +1$. The current noise can also be
expressed as $S_{in}^{1/2}=2\sqrt{eI_{B}}\left(  1+\Gamma_{\downarrow}%
/\Gamma_{\uparrow}\right)  ^{-1}$. In the limit of small $\Gamma_{\downarrow
}/\Gamma_{\uparrow}$ the Bloch oscillation sequences are short compared the
total length of the ''duty cycle'' $1/\Gamma_{\downarrow}+1/\Gamma_{\uparrow}%
$. Then the equivalent current noise can be understood to be simply the shot
noise of the input current. In that case $S_{in}^{1/2}=2\sqrt{eI_{B}}$. The
prefactor 2 instead of more familiar $\sqrt{2}$ is due to the random length of
charge pulses as opposed to standard shot noise. With large $\Gamma
_{\downarrow}/\Gamma_{\uparrow}$, or with long Cooper pair sequences, the
noise drops. The impedance also increases because single electron tunneling is
forbidden during the Bloch oscillations. One should remember, however, that
this is strictly true only in the absence of base junction leakage current.

Lowering the equivalent current noise below the input current shot noise level
can be understood as follows. In the limit governed by Eqs. (\ref{betaap0}%
)-(\ref{tnap0}) the length of the duty cycle is determined by the base current
$I_{B}$ (Eq. (\ref{Ibgen}) with $\left\langle N_{e}\right\rangle =1$). If we
have very short Cooper pair sequences (or short $1/\Gamma_{\uparrow}$)
compared to the duty cycle, the output current is essentially a sequence of
short charge pulses of size $2e\left\langle N\right\rangle .$ This leads to
shot type noise at the output, i.e. $S_{i,out}^{1/2}=2\sqrt{2e\left\langle
N\right\rangle I_{C}}$ (Fig. 6(a)). The equivalent noise at the input is then
$S_{i,in}^{1/2}\approx S_{i,out}^{1/2}/2\left\langle N\right\rangle $. Using
$I_{B}\approx I_{C}/2\left\langle N\right\rangle $ we then get $S_{i,in}%
^{1/2}=2\sqrt{eI_{B}}$. However, if the base current is constant and the
length of the Cooper pair sequence $1/\Gamma_{\uparrow}$ (or equivalently
$\left\langle N\right\rangle $) is increased by increasing the Josephson
coupling we eventually have a situation, where $1/\Gamma_{\downarrow}$ is the
same as $1/\Gamma_{\uparrow}$ was initially (see Fig. 6(b)). The current noise
at the output is obviously the same is both cases, but the current gain
$\ 2\left\langle N\right\rangle $ is larger in Fig. 6(b). Therefore the
equivalent noise\ at the input must be smaller.

\begin{center}
\begin{figure}[!b]
\includegraphics[width=6cm]{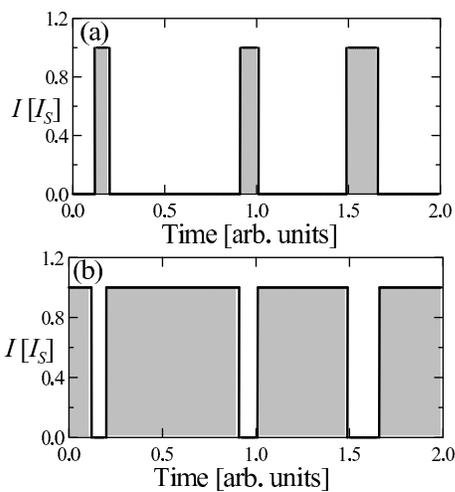}
\caption{Collector current $I_{C}$ schematically as function of time, when BOT is
operated in the limit, where $\left\langle N_{e}\right\rangle =1.$ In (a)
$\Gamma_{\downarrow}/\Gamma_{\uparrow}\ll1$ and (b) $\Gamma_{\downarrow
}/\Gamma_{\uparrow}\gg1$. The gray-shaded regions represent Cooper-pair
sequences with area $2e\left\langle N\right\rangle $.}
\end{figure}
\end{center}

As noted above, the spectral noise density of the backaction noise current
($i_{BA}$ in Fig. 5) in general differs from $S_{i,in}$. It can be shown, that
for either $\Gamma_{\uparrow}/\Gamma_{\downarrow}\ll1$ or $\Gamma_{\uparrow
}/\Gamma_{\downarrow}\gg1$, it is exactly\ that of the base current shot
noise, i.e. $\sqrt{2eI_{B}}$. The maximum suppression of $i_{BA} $ occurs at
$\Gamma_{\uparrow}=\Gamma_{\downarrow}$, where the fano factor is $1/2$. The
reason for the difference in the equivalent current noise and the backaction
noise is, that in the limit of large $\Gamma_{\downarrow}/\Gamma_{\uparrow}$
the output current noise becomes fully anticorrelated with $i_{BA}$. Thus
$i_{BA}$ does not directly determine the current resolution, or vice versa. To
minimize the backaction noise, the device should be operated at a low base
current. The low limit is here is set by spontaneous downwards transitions due
to incoherent Cooper pair tunneling \cite{del2}.

To investigate the parameters from the device optimization point of view, we
assume a typical point of operation, where $V_{C}=\left(  3/2\right)  V_{Q}$
and $V_{B}^{\prime}=-V_{Q}$. Then it follows%

\begin{align}
\beta &  \approx2\exp\left(  \frac{\pi Re^{2}}{8\hbar}\left(  \frac{E_{J}%
}{E_{C}}\right)  ^{2}\right) \label{betaap1}\\
g_{m} &  \approx-\frac{\beta}{R_{T}}\frac{1}{\left(  1+\left(  \ln5\right)
/2\left(  R_{T}/R\right)  \beta\right)  ^{2}}\label{gmap1}\\
S_{in}^{1/2} &  \approx\sqrt{\frac{4E_{C}}{R_{T}}\left(  1+\frac{\ln5}{2}%
\frac{R}{R_{T}}\beta\right)  ^{-3}}\label{Sinap1}\\
S_{en}^{1/2} &  \approx\sqrt{4E_{C}R_{T}\left(  1+\frac{\ln5}{2}\left(
\frac{R_{T}}{R}\right)  \beta\right)  }\label{Senap1}\\
R_{opt} &  \approx R_{T}\left(  1+\frac{\ln\left(  5\right)  }{2}\frac
{R}{R_{T}}\beta\right)  ^{2}\label{Roptap1}\\
T_{n} &  \approx\frac{E_{C}}{k_{B}}\left[  1+\frac{\ln\left(  5\right)  }%
{2}\left(  \frac{R}{R_{T}}\right)  \beta\right]  ^{-1}.\label{Tnap1}%
\end{align}
These formulas suggest, that Josephson coupling should be made large to
maximize the current gain and to minimize the added noise. One should also
remember that the fluctuation effects mentioned in Section IV and Ref.
\cite{del2} tend to suppress the gain which also increases the equivalent
noise. Also the assumption $E_{J}/E_{C}\lesssim1$ has to remain valid for the
model to work.

If the approximation from Eqs. (\ref{ne3}) and (\ref{gdown3}) is used instead
of Eqs. (\ref{ne1}) and (\ref{gdown2}) for calculating $\Gamma_{\downarrow}$
and $\left\langle N_{e}\right\rangle $, the dominating terms are in many cases
those dependent on $\beta_{B}$ especially if $\beta_{B}\approx1$. In Appendix
A it is presented a derivation of small signal and noise parameters.
Approximations for $\Gamma_{\uparrow}$, $\Gamma_{\downarrow}$, $\left\langle
N_{e}\right\rangle $, $\left\langle N\right\rangle $, $\partial\left\langle
N_{e}\right\rangle /\partial V_{B}$ and $\partial\Gamma_{\downarrow}/\partial
V_{B}$ are made by eliminating the bias parameters at an interesting point of
operation. ''Fine tuning'' of the device properties can be made by changing
$\beta_{B}$, which in our approximation stands%

\begin{equation}
\beta_{B}=0.02\left(  \frac{R}{R_{T}}\right)  ^{2}\exp\left(  \frac{\pi
e^{2}R}{16\hbar}\left(  \frac{E_{J}}{E_{c}}\right)  ^{2}\right)
,\label{betaBappr}%
\end{equation}
while other quantities of interest are%

\begin{align}
\beta &  \approx1.2\left(  1-\beta_{B}\right)  ^{-1}\label{betaap3}\\
g_{m} &  \approx-\frac{2}{R}\label{gmap3}\\
S_{in}^{1/2} &  \approx\frac{12e}{\sqrt{RC}}\left(  \frac{R_{T}}{R}\right)
\beta^{-1}\\
S_{en}^{1/2} &  \approx\frac{2e}{\sqrt{RC}}R_{T}\label{Sinap3}\\
R_{opt} &  \approx\frac{R}{2}\beta\label{Roptap3}\\
T_{n} &  \approx\frac{50E_{C}}{k_{B}}\left(  \frac{R_{T}}{R}\right)  ^{2}%
\beta^{-1}\label{Tnap3}%
\end{align}
As $\beta_{B}\rightarrow1$ the current gain $\beta$ diverges. However, the
trade-off is that the optimum impedance $R_{opt}$ also diverges. The
fluctuation at the output does not depend on $\beta_{B}$, so the current noise
$S_{in}^{1/2}$ and the noise temperature $T_{n}$ decrease at the same time.
The transconductance gain $g_{m}$ and voltage noise $S_{en}^{1/2}$ are
independent of $\beta_{B}$.

\begin{center}
\begin{figure}[!t]
\includegraphics[width=8cm]{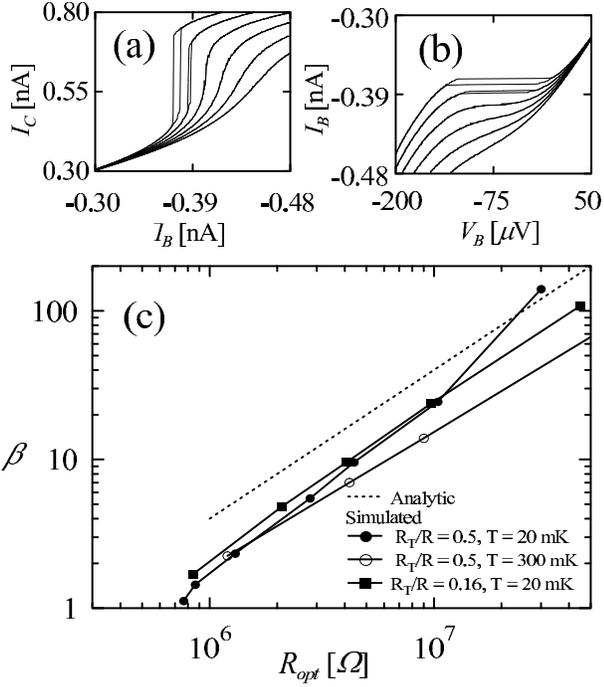}
\caption{(a) and (b) show a computed set ($R=500$ k$\Omega$, $C=1.2$ fF,
$\Delta=400$ $\mu$V, $T=20$ mK and $R_{T}/R=0.5$) of characteristic curves
used to extract $\beta$ and $R_{opt}$. The Josephson coupling is varied from
$E_{J}/E_{C}=0.18$ to $E_{J}/E_{C}=0.28$ (from right to left in (a) and down
to up in (b). (c)\ Computed and analytic current gains $\beta$ plotted against
the optimum resistance $R_{opt}$ as the Josephson coupling (or $\beta_{B}$) of
the device is varied. The parameters are as above with the exceptions of
varying $R_{T}/R$ and $T$ as shown in the legend. The bias point in the
simulations with $T=20$ mK is $V_{C}=3.5e/C$ and $V_{C}=4.5e/C$ for those with
$T=300$ mK.}
\end{figure}
\end{center}

The physics in this limit can be understood as follows. With very large
$\beta_{B}$ the main effect of increasing $V_{B}$ is increasing the number of
electrons $\left\langle N_{e}\right\rangle $ needed to cause a downwards
transition (see Eq. \ref{betabdef}). This leads to decreasing $I_{B}$, i.e.
negative input conductance. With very small $\beta_{B}$ the only effect of
increasing $V_{B}$ is decreasing $\Gamma_{\downarrow}$. This leads to
increasing $I_{B}$, i.e. positive input conductance \cite{exp1}. At
intermediate values, i.e. $\beta_{B}\approx1$, the input conductance is close
to zero. The effect is that a small change in $I_{B}$ causes a large change in
$V_{B}$. Consequently $\Gamma_{\downarrow}$, and thus also $I_{C} $ change
considerably. This leads to the enhancement of the current gain.

A set of simulated $I_{C}-I_{B}$ and $I_{B}-V_{B}$-plots with a varying
Josephson coupling are shown in Figs. 7(a) and 7(b). The parameters were
chosen so that the device is realizable with Al-tunnel junctions (see the
Caption of Fig. 7). Current biased base electrode was assumed. This shows how
the current gain and the input impedance increase without limit, as $\beta
_{B}$ approaches unity. As $\beta_{B}$ exceeds unity the curves become
hysteretic. If the source resistance $R_{S}$ is large, hysteresis is a
manifestation of negative input conductance. Therefore a sufficient stability
criterion for all source resistances is $\beta_{B}<1$. For small source
resistances the device is stable independently of $\beta_{B}$. The simulated
IV curves become hysteretic at $E_{J}/E_{C}\approx0.25.$ According to Eq.
(\ref{ne3}) $E_{J}/E_{C}\approx0.32$ leads to $\beta_{B}=1 $.

We will next illustrate the trade-offs and test the validity of Eqs.
(\ref{betaBappr})-(\ref{Tnap3}). The computed current gain as function of
optimum resistance is shown together with the analytic approximation obtained
from Eqs. (\ref{betaap3}) and (\ref{Roptap3}) in Fig. 7(c). Each of the three
sets have different base resistance $R_{T}$ and the bath temperature. Within
each set $E_{J}/E_{C}$ is varied. One can see, that the dependence
$\beta\propto R_{opt}$ is correctly reproduced regardless of parameters, i.e.
the property is quite generic.

The current noise and the minimum noise temperature are shown as the function
of the optimum resistance in Fig. 8. The computational noise data was obtained
by performing a Fast Fourier Transform for the output current and averaging
the low-frequency part, which gives $S_{i,out}$. To further evaluate $S_{in}$,
$R_{opt}$ and $T_{n}$, Eqs. (\ref{Sinout}), (\ref{Roptdef}) and (\ref{Tndef})
were applied with computed current gain $\beta$ and input resistance $R_{in}$.
Again the correct form of dependencies, i.e. $S_{in}^{1/2}\propto R_{opt}%
^{-1}$ and $T_{n}\propto R_{opt}^{-1}$ are correctly reproduced as compared to
Eqs. (\ref{Roptap3}) and (\ref{Tnap3}). Differences in absolute levels can
partially be explained through the inaccuracy of the numeric constants due to
approximations made in Appendix A. To some extent the differences can also be
understood with reference to excess noise mechanisms discussed in Section VII.
However, correct forms of dependencies and the order of magnitude are
correctly predicted with the theory.

\begin{center}
\begin{figure}[!t]
\includegraphics[width=8cm]{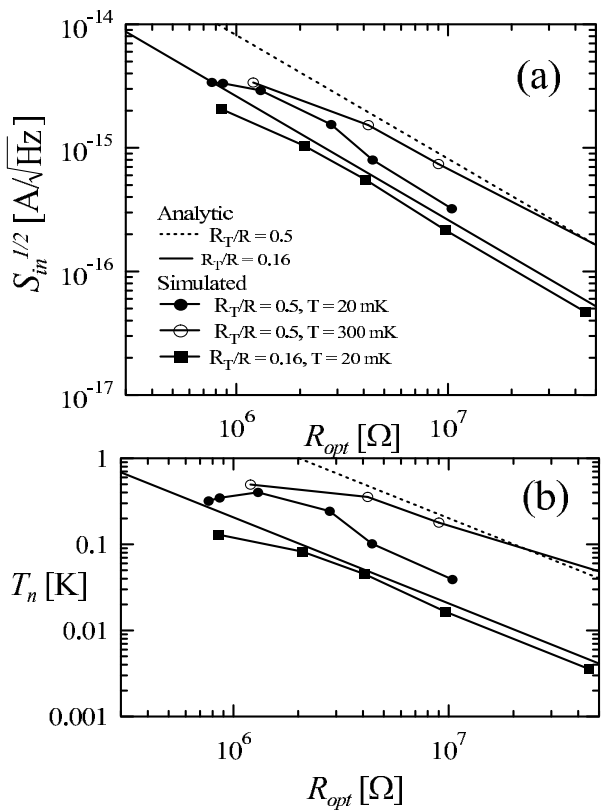}
\caption{(a)\ Computed and analytic results for the current noise spectral density
$S_{in}$ referred to input and (b)\ the minimum noise temperature $T_{n}$ as
function of $R_{opt}$. The device and bias parameters are the same as in 
Fig. 7.}
\end{figure}
\end{center}

\section{Summary and discussion}

We have developed an analytic theory to predict characteristic curves and
noise properties of the Bloch Oscillating Transistor. Even though it was
derived at zero temperature, comparison to simulations at finite temperatures
showed reasonable agreement. The reason is that the main fluctuation
mechanism, the two-level switching noise, is essentially temperature
independent. Of other noise sources, for example, the thermal noise of the
base junction is insignificant, since the junction is at a typical point of
operation biased at $eV_{1}\gg kT$.

Two modes of operation were discussed. The small signal and noise parameters
in both limits were obtained, i.e. Eqs. (\ref{betaap1})-(\ref{Tnap1}) and Eqs.
(\ref{betaap3})-(\ref{Tnap3}). In the first mode the device acts as a simple
$e-2e\left\langle N\right\rangle $ charge multiplier. In the second mode
intralevel transitions play an important role. The consequence is the
emergence of the hysteresis parameter $\beta_{B}$. This was found to have a
drastic effect on device properties. It was shown that noise currents below 1
fA and noise temperatures below 100 mK can be obtained for optimum impedances
of order a few M$\Omega$.

An additional noise mechanism is the shot-noise of the leakage current from
the base-electrode. At a finite temperature also the bandwidth of the
Bloch-oscillation increases\cite{lik1}, and thus the shot-noise of the Cooper
pairs adds to the total noise. This may explain the factor of about 3 increase
in noise temperature as temperature is changed from 20 mK to 300 mK in Fig.
8(b). These effects appear at the output of the amplifier. Thus they add to
the total output noise being additional terms to Eq. (\ref{Sout}). Even in the
presence of them the main conclusions of this article remain valid.

We have also assumed arbitrarily large $C_{B}$, which is acceptable in
low-frequency applications. However, if one wishes to increase the band,
$C_{B}$ should be decreased. The first effect of finite $C_{B}$ is that the
base voltage starts to fluctuate at frequencies typical to BOT\ dynamics.

Most other well-known mesoscopic amplifiers, e.g. single-electron transistor
(SET) \cite{ave1} or single Cooper pair transistor (SCPT) \cite{zor1} are
based on controlling current flow by charging a gate electrode. This is
similar to the field effect transistor (FET), whereas BOT\ resembles a bipolar
junction transistor (BJT)\ in the sense that a small base current is used to
control a larger collector current. However, there are also important
differences as well. For example, we have shown that the equivalent current
noise of BOT\ can be brought below the shot noise of the input current. The
reason is, that in BOT\ the noise at the output is partially correlated to the
noise at the input.

The $1/f$-noise of the BOT\ is not addressed in this paper. However, as
opposed to gate-controlled devices, the BOT is immune to background charge
fluctuations. It is probable that the main $1/f$-noise mechanism is the
fluctuation of the Josephson coupling. Due to symmetry considerations, it can
be reduced by using bias reversal techniques typically used with a dc SQUID
(\cite{dru1}).
\\
Authors wish to acknowledge fruitful discussions with P. Hakonen and R.
Lindell. The work was supported by the Academy of Finland (project 103948).

\appendix
\section{Derivation of hysteresis and noise parameters}

In this Appendix we show the derivation of Eqs. (\ref{betaBappr}%
)-(\ref{Tnap3}). The aim is to derive an approximation, which applies at
$\beta_{B}\approx1$ and $R_{T}\lesssim R$.

The main task is to get sufficient estimates for $I_{S},$ $\Gamma_{\downarrow
}$, $\Gamma_{\uparrow}$, $\left\langle N_{e}\right\rangle $, $\partial
\left\langle N_{e}\right\rangle /\partial V_{B}$ and $\partial\Gamma
_{\downarrow}/\partial V_{B}$ near an interesting point of operation. We note
that for the approximation from Eqs. (\ref{ne3}) and (\ref{gdown3}) used for
$\left\langle N_{e}\right\rangle $ and $\Gamma_{\downarrow}$ to apply the
collector voltage must satisfy $V_{C}\gtrsim2V_{Q}$. For simplicity we assume
that $V_{C}=2V_{Q}$. When the system is at the second level, the voltage
across the JJ is $V_{2}\gtrsim V_{Q}$ and the collector current consists of
the leakage current only, i.e. $I_{C}=-I_{B}$. Assuming that $V_{2}=V_{Q}$ and
that the base junction roughly acts as a linear resistor we get by analyzing
the circuit of Fig. 1(a) and noting that by definition $V_{B}=V_{B}^{\prime
}+V_{C}$ the result $V_{B}^{\prime}=-V_{Q}\left(  1+R_{T}/R\right)  $, i.e. we
have found estimates for the bias parameters. Using these and Eqs. (\ref{Is}),
(\ref{Gup}) and (\ref{ne3}) we can readily write:%

\begin{align}
I_{S}  &  \approx\frac{2e}{RC\ln3}\label{Isap}\\
\Gamma_{\uparrow}  &  \approx\frac{1}{RC\ln3}\exp\left(  -\frac{\pi RC}%
{8\hbar}\frac{E_{J}^{2}}{E_{c}^{2}}E_{C}\right) \label{Gupap}\\
\left\langle N_{e}\right\rangle  &  \approx100\left(  \frac{R_{T}}{R}\right)
^{2}.\label{Neap}%
\end{align}
In the last Equation we have also utilized the assumption $R_{T}\lesssim R$.

To find an estimate for $\Gamma_{\downarrow}$ we can use physical intuition
and insight learned from simulations. Since the operation is based on
switching between the two states, the system spends roughly as much time in
both states. Thus

$\bigskip$%
\begin{equation}
\Gamma_{\downarrow}\approx\Gamma_{\uparrow}\text{.}\label{Gdownap}%
\end{equation}
Although some error may be introduced by doing this, it is not too severe,
since most properties depend more strongly on the derivative $\partial
\Gamma_{\downarrow}/\partial V_{B}$, which will be calculated separately.

The derivative $\partial\left\langle N_{e}\right\rangle /\partial V_{B}$ is
obtained by direct differentiation of Eq. (\ref{ne3}), inserting the bias
parameters $V_{B}$, $V_{C}$ from above and applying the approximation
$R_{T}\lesssim R$:%

\begin{equation}
\frac{\partial\left\langle N_{e}\right\rangle }{\partial V_{B}}\approx
\frac{270}{V_{Q}}\left(  \frac{R_{T}}{R}\right)  ^{2}.\label{dNeap}%
\end{equation}

To obtain $\partial\Gamma_{\downarrow}/\partial V_{B}$ we first note that
$\Gamma_{\downarrow}$ depends on $V_{B}$ both through the explicit dependence
in Eq. (\ref{gdown3}) and through $\left\langle N_{e}\right\rangle $.\ Since
$\left\langle N_{e}\right\rangle $ depends very strongly on $V_{B}$ we will
approximate $\partial\Gamma_{\downarrow}/\partial V_{B}\approx\left(
\partial\Gamma_{\downarrow}/\partial\left\langle N_{e}\right\rangle \right)
\left(  \partial\left\langle N_{e}\right\rangle /\partial V_{B}\right)  $. By
differentiation of Eq. (\ref{gdown3}), application of the bias parameters
$V_{B}$ and $V_{C}$ from above and using Eq. (\ref{dNeap}) we get%

\[
\frac{\partial\Gamma_{\downarrow}}{\partial V_{B}}\approx1.2e\Gamma
_{\downarrow}^{2}\frac{R+R_{T}}{V_{B}^{\prime}}\frac{\partial\left\langle
N_{e}\right\rangle }{\partial V_{B}}.
\]
By using Eqs. (\ref{Gdownap}) and (\ref{Gupap}) we now get%

\begin{equation}
\frac{\partial\Gamma_{\downarrow}}{\partial V_{B}}\approx-\frac{290}{V_{Q}%
}\frac{1}{RC}\left(  \frac{R_{T}}{R}\right)  ^{2}\left(  \exp\left(
-\frac{\pi RC}{8\hbar}\frac{E_{J}^{2}}{E_{c}^{2}}E_{C}\right)  \right)
^{2}.\label{dGdownap1}%
\end{equation}
\qquad\ 

By inserting Eqs. (\ref{Isap})-(\ref{dGdownap1}) into Eq. (\ref{betabdef}) we
get the estimate of the hysteresis parameter $\beta_{B}$, which is shown in
Eq. (\ref{betaBappr}).

The exponent (or $\left\langle N\right\rangle $) in Eqs. (\ref{Gupap}) and
(\ref{dGdownap1}) affects on the device parameters mainly through $\beta_{B}$.
For other purposes we may assume it roughly constant and solve it by setting
$\beta_{B}=1$ in Eq. (\ref{betaBappr}), whence Eqs. (\ref{Gupap}) and
(\ref{dGdownap1}) are simplified into%

\begin{align}
\Gamma_{\uparrow}  &  \approx\frac{0.015}{RC}\left(  \frac{R}{R_{T}}\right)
^{2}\label{Gupap2}\\
\frac{\partial\Gamma_{\downarrow}}{\partial V_{B}}  &  \approx-\frac
{0.08}{V_{Q}}\frac{1}{RC}\left(  \frac{R}{R_{T}}\right)  ^{2}%
.\label{dGdownap2}%
\end{align}

Eqs. (\ref{gmap3})-(\ref{Tnap3}) are now calculated by substituting Eqs.
(\ref{Isap}), (\ref{Neap}), (\ref{Gdownap}), (\ref{dNeap}), (\ref{Gupap2}) and
(\ref{dGdownap2}) into the definitions of interesting quantities, i.e. Eqs.
(\ref{gmgen}), (\ref{betagen}), (\ref{Sen}), (\ref{Sin}), (\ref{Roptdef}) and
(\ref{Tndef}).


\begin{thebibliography}{99}
\bibitem {exp1}Note that with our sign conventions ($V_{B}<0$, $I_{B}<0$)
electrons are tunneling to the island, i.e. Increasing $V_{B}$ decreases
$\Gamma_{\downarrow}$, which increases $I_{B}$. On the other hand, increasing
$V_{B}$ increases $\left\langle N_{e}\right\rangle $, which decreases $I_{B}$.
\end{thebibliography}
\end{document}